\documentstyle[pasjskel,graphicx,natbib,twocolumn]{article}

\begin{document}

\title{The Extended Deep Minimum and the Subsequent Brightening of RX~And
       in 1996--1997}

\author{Taichi \textsc{Kato}}
\affil{Department of Astronomy, Kyoto University,
       Sakyo-ku, Kyoto 606-8502}
\email{tkato@kusastro.kyoto-u.ac.jp}

\author{Daisaku \textsc{Nogami}}
\affil{Hida Observatory, Kyoto University, Kamitakara, Gifu 506-1314}
\email{nogami@kwasan.kyoto-u.ac.jp}

\email{\rm{and}}

\author{Seiji \textsc{Masuda}}
\affil{Okayama Astrophysical Observatory, National Astronomical
       Observatory, Okayama 719-0232}


\begin{abstract}
   We discovered that RX And, one of prototypical Z Cam-type dwarf novae,
underwent a deep, extended faint state in 1996--1997.  Time-resolved
photometry at the bottom of the fading revealed the presence of strong
flickering, and the absence of detectable orbital modulation.  This
finding indicates that the mass-transfer remained even at the deepest
minimum of the fading, contrary to what was observed in a deep minimum
of a VY Scl-type star, MV Lyr.  RX And subsequently underwent a brightening
(outburst) during its recovery stage.  The photometric and spectroscopic
characteristics of the brightening significantly differed from those of
ordinary outbursts of RX And, and are considered to resemble an
inside-out type outburst of a long-period dwarf nova.
An examination of historical visual observations revealed the possible
presence of $\sim$10-yr periodicity, which is close to what has been
proposed for MV Lyr.  The common observed features between RX And and
VY Scl-type stars may suggest a common underlying mechanism for producing
temporary deep fadings.  The departure from the disk instability model,
as observed in VY Scl-type stars, was not apparent in the present
fading of RX And.  In conjunction with the recently published Hubble Space
Telescope observation during the same fading, we can conclude that the
phenomenological difference from the VY Scl-type fading is understood
as a smaller effect of irradiation on the accretion disk in RX And.
\end{abstract}

KeyWords: accretion, accretion disks
          --- stars: binaries: eclipsing
          --- stars: dwarf novae
          --- stars: novae, cataclysmic variables
          --- stars: individual (RX Andromedae)

\section{Introduction}

   RX And is one of the best-known Z Cam-type dwarf novae (cf.
\cite{war95book}), which show standstills in addition to usual dwarf
nova-type outbursts (for a review of dwarf nova outbursts,
see \cite{osa96review}).  Although the exact origin of Z Cam-type
phenomenon is not perfectly understood, it has been proposed that
the changing mass-transfer rate ($\dot{M}$) is most responsible for
the dwarf nova--standstill alternations: the standstill is a state
of an enhanced ($\dot{M}$), which thermally stabilizes the accretion
disk (e.g. \cite{mey83zcam}).  \citet{mey83zcam} proposed that
a normal outburst below the critical surface density can
trigger a standstill, which is maintained by an enhanced mass-transfer
caused by irradiation from the brightened accretion disk and white
dwarf.  Subsequent model calculations by \citet{lin85CValphadisk} and
\citet{kin98DI}, however, did not well reproduce the observed properties
of Z Cam-type stars.  \citet{hon98zcam} presented an observational test,
and reported that some Z Cam stars were indeed brighter in standstill
than the mean brightness in the outbursting phase, suggesting higher
mass-transfer rates during standstills.

   More recently, \citet{bua01zcam} claimed that the basic features are
reproduced by correctly taking the heating of the disk by the tidal torque
and the mass-transfer stream impact into account.  \citet{bua01zcam}
also claimed that large variation of mass-transfer rates are not necessary
to reproduce the Z Cam-type phenomenon.  However, the resultant light
curve by \citet{bua01zcam} is different from the observation in some
respects.  \citet{kat01zcam} argued that there is a striking departure
from the theoretical prediction when a system enters a standstill.

   Several types of cataclysmic variables (CVs) show temporary reductions
of $\dot{M}$, the best-known cases being VY Scl-type stars and AM Her-type
stars (polars) (\cite{war95book}; \cite{gre98vyscl}).
Indeed, \citet{war74vysclzcam} tried to explain the Z Cam-type and
VY Scl-type phenomena in the same framework.
However, observations suggest that there is a strong segregation of
period distributions between Z Cam-type stars and VY Scl-type stars
\citep{ver97Porb}.  Furthermore, past extensive studies on representative
Z Cam-type stars (\cite{hon98zcam}; \cite{ope98zcam}), together with
more theoretical discussion by \citet{bua01zcam}, have shown little
evidence for VY Scl-type strong temporary reduction of $\dot{M}$.

   The nature of Z Cam-type phenomenon is thus still controversial;
there still remain discrepancies between observations and theories.
Even if we assume changing $\dot{M}$ as a cause of Z Cam-type
phenomenon, the exact origin of changing $\dot{M}$ is still poorly
understood, although star spots covering the L$_1$ point
\citep{liv94CVstarspot} would be a promising interpretation for
the VY Scl-type phenomena.

  RX And had been known as a relatively typical Z Cam-type dwarf nova,
with outbursts-standstill alternations, until the discovery of an
extremely deep fading (also called deep quiescence by some authors)
in 1996 September.\footnote{
  VSNET observations, No. 3750,
  http://www.kusastro.kyoto-u.ac.jp/vsnet/Mail/obs3000/msg00750.html
}
The deep fading lasted until 1997 January.

\section{Observation}

\subsection{Photometry}

   The time-resolved observations were acquired on three nights between 1996
November 15 and 19 (near the bottom of the fading), using a CCD camera
(Thomson TH~7882, 576 $\times$ 384 pixels, on-chip 2 $\times$ 2 binning
adopted) attached to the Cassegrain focus of the 60 cm reflector
(focal length=4.8 m) at Ouda Station, Kyoto University \citep{Ouda}.
An interference filter was used which had been designed to reproduce the
Johnson $V$ band.  The exposure time was 60 s.  The frames were first
corrected for standard de-biasing and flat fielding, and were then
processed by a microcomputer-based aperture photometry package developed
by one of the authors (TK).  The magnitudes of the object was determined
relative to GSC 2807.1483 ($V$=11.86, $B-V$=+1.05), whose constancy
during the run was confirmed using GSC 2803.948.  The magnitude of the
comparison star is taken from \citet{mis96sequence}.
Barycentric corrections to observed times were applied before the
following analysis.  We also obtained additional nightly snapshot
photometry on 32 nights between 1996 September 10 and 1997 February 5.

Table \ref{tab:log} lists the log of observations, together with nightly
averaged magnitudes.

\begin{table}
\begin{center}
\caption{Nightly averaged magnitudes of RX And}\label{tab:log}
\begin{tabular}{lcrccccr}
\hline\hline
\multicolumn{3}{c}{Date} &
  Start$^*$ & End$^*$ & mean$^\dagger$ & err$^\ddagger$ & N$^\S$ \\
\hline
1996 & Sep & 10 & 337.265 & 337.266 &  2.802 & 0.027 &   3 \\
1996 & Sep & 15 & 342.175 & 342.176 &  3.146 & 0.025 &   3 \\
1996 & Sep & 16 & 343.124 & 343.125 &  2.929 & 0.039 &   3 \\
1996 & Sep & 17 & 344.099 & 344.100 &  3.333 & 0.036 &   3 \\
1996 & Sep & 18 & 345.040 & 345.041 &  2.949 & 0.071 &   3 \\
1996 & Oct & 20 & 377.141 & 377.144 &  3.313 & 0.028 &   5 \\
1996 & Oct & 22 & 379.154 & 379.157 &  3.394 & 0.022 &   5 \\
1996 & Oct & 29 & 386.204 & 386.210 &  3.290 & 0.017 &   8 \\
1996 & Nov & 15 & 402.883 & 403.135 &  3.294 & 0.007 & 265 \\
1996 & Nov & 16 & 403.899 & 404.200 &  3.304 & 0.006 & 320 \\
1996 & Nov & 19 & 406.982 & 407.204 &  3.272 & 0.005 & 280 \\
1996 & Dec &  8 & 425.870 & 425.875 &  2.825 & 0.045 &   4 \\
1996 & Dec &  9 & 426.938 & 426.943 &  3.219 & 0.014 &   5 \\
1996 & Dec & 11 & 428.896 & 428.898 &  3.315 & 0.022 &   5 \\
1996 & Dec & 18 & 435.930 & 435.933 &  3.091 & 0.092 &   5 \\
1996 & Dec & 21 & 439.010 & 439.011 &  3.346 & 0.140 &   3 \\
1996 & Dec & 22 & 439.870 & 439.871 &  3.231 & 0.053 &   3 \\
1996 & Dec & 23 & 440.858 & 440.858 &  2.908 & 0.173 &   2 \\
1996 & Dec & 24 & 441.860 & 441.861 &  2.955 & 0.033 &   3 \\
1996 & Dec & 25 & 443.007 & 443.008 &  3.267 & 0.056 &   3 \\
1996 & Dec & 27 & 444.868 & 444.873 &  3.066 & 0.019 &   5 \\
1996 & Dec & 31 & 448.865 & 448.866 &  3.067 & 0.168 &   3 \\
1997 & Jan &  1 & 449.867 & 449.869 &  2.895 & 0.142 &   3 \\
1997 & Jan &  2 & 450.912 & 450.918 &  2.763 & 0.017 &  12 \\
1997 & Jan &  3 & 451.869 & 451.870 &  2.571 & 0.019 &   3 \\
1997 & Jan &  4 & 452.927 & 452.929 &  2.185 & 0.013 &   5 \\
1997 & Jan &  6 & 454.867 & 545.871 &  2.607 & 0.073 &   8 \\
1997 & Jan &  8 & 456.865 & 456.869 &  1.296 & 0.015 &   6 \\
1997 & Jan & 12 & 460.930 & 460.932 &  0.020 & 0.002 &   5 \\
1997 & Jan & 13 & 461.867 & 461.869 & -0.333 & 0.004 &   5 \\
1997 & Jan & 15 & 463.865 & 463.870 & -0.506 & 0.004 &  10 \\
1997 & Jan & 19 & 467.907 & 467.909 &  1.537 & 0.008 &   5 \\
1997 & Jan & 20 & 468.875 & 468.877 &  1.926 & 0.015 &   5 \\
1997 & Feb &  3 & 482.906 & 482.909 &  0.813 & 0.005 &   5 \\
1997 & Feb &  5 & 484.906 & 484.911 &  1.736 & 0.013 &   6 \\
\hline
  \multicolumn{8}{l}{$^*$ BJD$-$2450000.} \\
  \multicolumn{8}{l}{$^\dagger$ Relative magnitude to GSC 2807.1483.} \\
  \multicolumn{8}{l}{$^\ddagger$ Standard error of nightly average.} \\
  \multicolumn{8}{l}{$^\S$ Number of frames.} \\
\end{tabular}
\end{center}
\end{table}

\begin{figure}
  \begin{center}
    \FigureFile(80mm,60mm){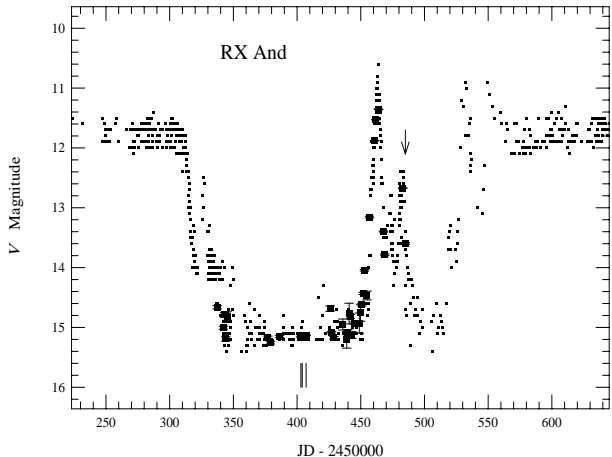}
  \end{center}
  \caption{Light curve of the extended faint state of RX And occurring
  in 1996.  The light curve is drawn from visual observations reported
  to VSNET (dots) and present CCD $V$-band observations (filled squares
  with error bars).  The vertical ticks and arrow represent
  the epochs of time-resolved photometry and spectroscopy, respectively.}
  \label{fig:overall}
\end{figure}

\subsection{Spectroscopy}

We took a low-resolution spectrum with a 1.88-m telescope and the
New Cassegrain Spectrograph equipped with a 150 grooves mm$^{-1}$
at Okayama Astrophysical Observatory (OAO, a branch of the National
Astronomical Observatory, an inter-university research institute
operated by the Ministry of Education, Science, Sports and Culture of
Japan) on 1997 February 5.  The detector was a CI502AB CCD
(512$\times$512 pixels).  The spectral coverage was 4600--7100 \AA,
and the wavelength resolution was 5.9 \AA pix$^{-1}$.  The exposure
started at BJD 2450484.9111, and the exposure time was 600 s.

The reduction was done using the Spectronebulagraph reduction system
(SNGRED) developed mainly by M. Yoshida at OAO on IRAF
package\footnote{IRAF is distributed by the National Optical Astronomy
Observatories for Research in Astronomy, Inc. under cooperative
agreement with the National Science Foundation.}.  The averaged
signal-to-noise ratio ($S/N$) per pixel at the continuum level was 22.
The spectrum was normalized by the continuum level.

\section{Result and Discussion}

\subsection{The Fading Episode in 1996--1997}

   The overall light curve, which is drawn from the present CCD observations
and the visual observations reported to VSNET international variable star
network\footnote{
   http://www.kusastro.kyoto-u.ac.jp/vsnet/
}, is presented in figure \ref{fig:overall}.  For comparison with the
$V$=15 mag reached during this extended low state, note that RX And is
normally at $V$=14 in quiescence.

   Such a fading strongly resembles faint (or low) states observed in
VY Scl-type novalike variables.  The existence of such an extended,
extremely faint state in a Z Cam-type star firmly established
that a remarkable reduction of mass-transfer rates comparable to those
of VY Scl-type stars indeed occurs in a Z Cam-type star.
The faint state lasted for $\sim$140 d.

\subsection{Short-Term Light Variation}

   In some VY Scl-type stars, a total cessation of mass-transfer is
suggested from the disappearance of flickering (MV Lyr: \cite{rob81mvlyr}).
We obtained time-resolved photometry of RX And during this extremely
low state, in order to see whether or not the mass-transfer stopped.

\begin{figure}
  \begin{center}
    \FigureFile(88mm,120mm){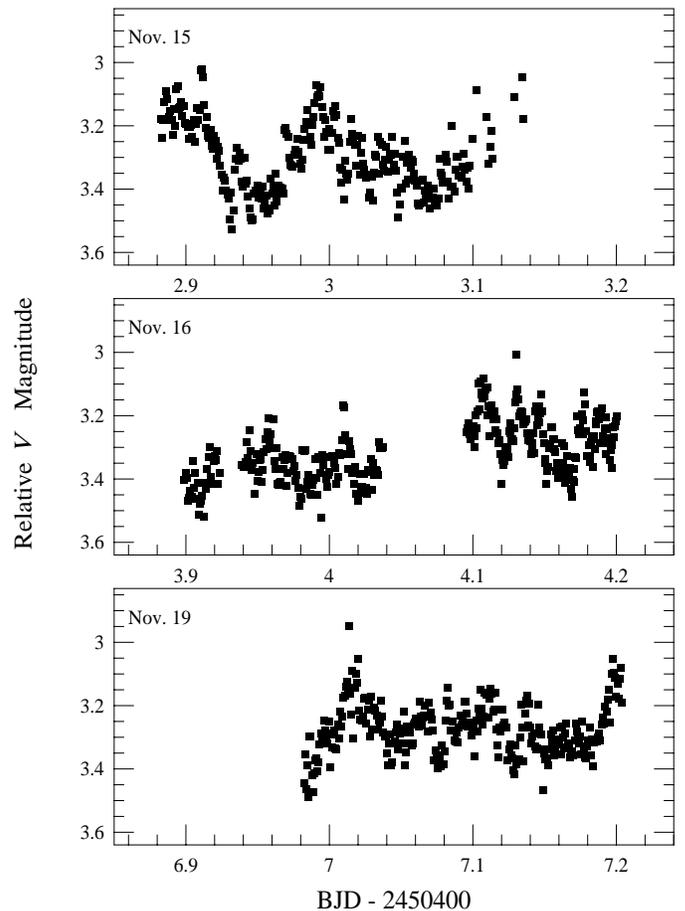}
  \end{center}
  \caption{Time-resolved photometry of RX And during the deep fading in
  1996.  The error of individual measurements are smaller than 0.01 mag.
  Short-term irregular variations are clearly visible, which can be
  attributed to flickering.}
  \label{fig:lc}
\end{figure}

\begin{figure}
  \begin{center}
    \FigureFile(88mm,60mm){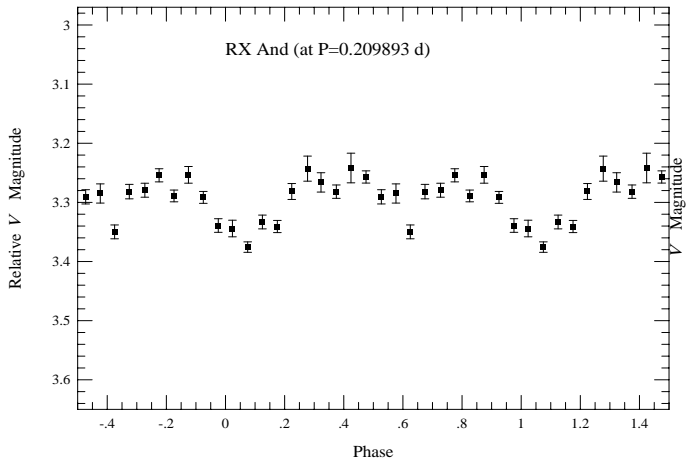}
  \end{center}
  \caption{Phase-averaged light curve of RX And during the deep fading.
  The upper limit of the amplitude of orbital modulations is 0.1 mag.
  }
  \label{fig:phase}
\end{figure}

   Figure \ref{fig:lc} shows the nightly light curves.  There indeed existed
short-term irregular variations, which can be attributed to flickering.
A period analysis with the Phase Dispersion Minimization (PDM) method
\citep{PDM} or discrete Fourier transform did not yield any
coherent periodicity.  The lack of coherent periodicity is also
evident in figure \ref{fig:lc} in that major flickerings (flare-like
brightenings) occurred randomly.  The variations were thus regarded as
irregular.  The variations were too irregular to be considered as
oscillations of the white dwarf (e.g. \cite{vanzyl01gwlib}).
The persistence of flickering indicates that a substantial
amount of mass-transfer in RX And was present, even during the extended
deep minimum.
Figure \ref{fig:phase} shows the phase-averaged light curve, at the
orbital period ($P$=0.209893 d, \cite{kai89rxand}).  Although there is an
indication of double-humped orbital variation with an amplitude of
$\sim$0.1 mag, the lack of coherent signal in the period analysis more
strongly suggests that the variation in the averaged light curve is
spurious.  We therefore put the upper limit of the amplitude of orbital
modulations as 0.1 mag.

One might expect a direct reflection effect from the hot white dwarf,
if the accretion disk had disappeared or become optically very thin.
The absence of such an effect also supports that a substantial amount of
accretion disk remained even in the extended deep minimum of RX And.
Future optical spectroscopic observations on similar occasions will be
able to test the presence of a normal CV disk.

\citet{sio01rxand} independently obtained far-UV spectra during the same
deep fading with the Hubble Space Telescope.  \citet{sio01rxand}
applied a synthetic spectral fitting to the UV spectra, and concluded
that the accretion disk contributes to 25 \% of the far-UV light,
corresponding to a mass-accretion rate of 10$^{-10.52}$ $M$$_{\odot}$
yr$^{-1}$.  Although their estimate of the accretion rate may have suffered
from some degree of ambiguities resulting from a fit using an optically
thick disk, the unmistakable signature of the remaining mass accretion
(close to the white dwarf) in the far-UV wavelengths strengthens our
finding of the remaining {\it mass-transfer} from the secondary.

\subsection{Brightening from the Minimum}

\begin{figure}
  \begin{center}
    \FigureFile(88mm,60mm){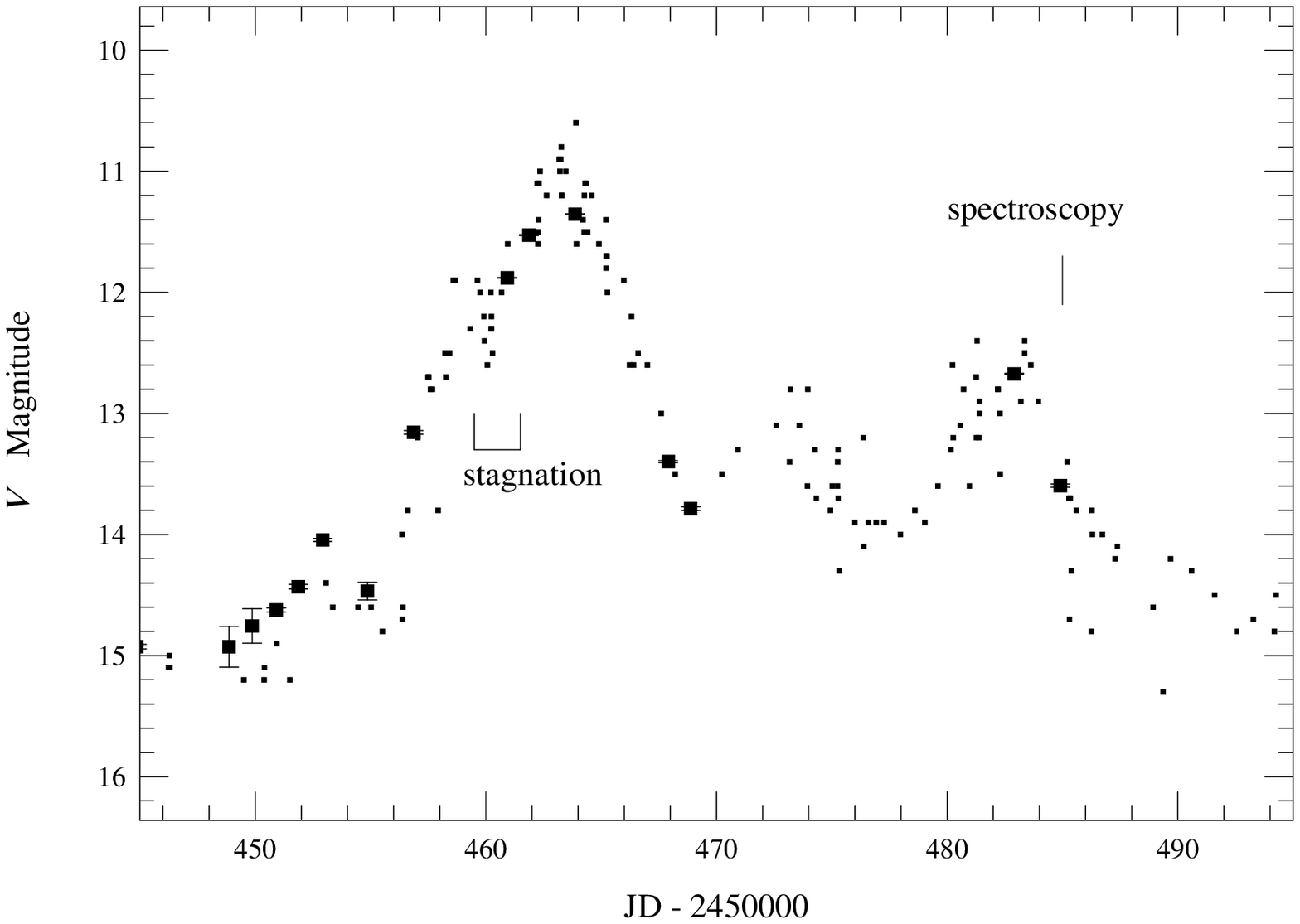}
  \end{center}
  \caption{Brightening from the deep fading.  The entire duration of the 
  brightening (outburst) was $\sim$40 d, with multiple peaks.  The rise to
  the maximum was anomalously slow, accompanied by a short ``stagnation"
  period in the rising branch.}
  \label{fig:brighten}
\end{figure}

   There was a gradually brightening trend since JD 2450439 (1996 December
21).  RX And brightened for the subsequent 14 nights at a mean rate of
0.037 mag d$^{-1}$.  Such a slow brightening is commonly observed in
the early recovery stage from deep, long, fadings of VY Scl-type stars
(cf. general light curves: \cite{gre98vyscl}; MV Lyr: \cite{kra92mvlyr}).
RX And then started brighten quickly (figure \ref{fig:brighten}),
attaining the maximum of $V$=11.4 on 1997 January 14.

   The light curve of the brightening more or less resembles that
of a dwarf nova-type outburst, but is different from ordinary outbursts of
RX And in many respects: 1) The entire duration of the brightening
(outburst) was $\sim$40 d, with multiple peaks.  The duration is several
times longer than those of ordinary outbursts of RX And
(\cite{szk74rxand}; \cite{szk84AAVSO}).  2) The rise to
the maximum was anomalously slow, accompanied by a short ``stagnation"
period in the rising branch (a temporary fading was particularly
apparent near JD 2450460 = 1997 January 11).  The duration of
the ``outburst" was similar to dwarf nova-like brightenings observed
in a VY Scl-type star, MV Lyr (\cite{pav98mvlyr}; \cite{shu98mvlyr};
\cite{pav99mvlyr}).

   The slow rising branch also resembles those of long-period dwarf
novae (\cite{men86bvcen}; \cite{kim92gkper}; \cite{sim00dxand}).
\citet{kim92gkper} interpreted this feature in GK Per as a result of
slow inside-out propagation of the thermal instability starting at the
inner region of the accretion disk.  This condition is apparently achieved
in long-period, low-$\dot{M}$ dwarf novae, in which the critical surface
density is more likely to be reached at the inner region \citep{kim92gkper}.

    The same condition was likely to be achieved in the recovery process
of RX And from the very low state, when the mass-transfer seems to have
slowly recovered, as inferred from the slow rise from the deep minimum.

\begin{figure}
  \begin{center}
    \FigureFile(88mm,60mm){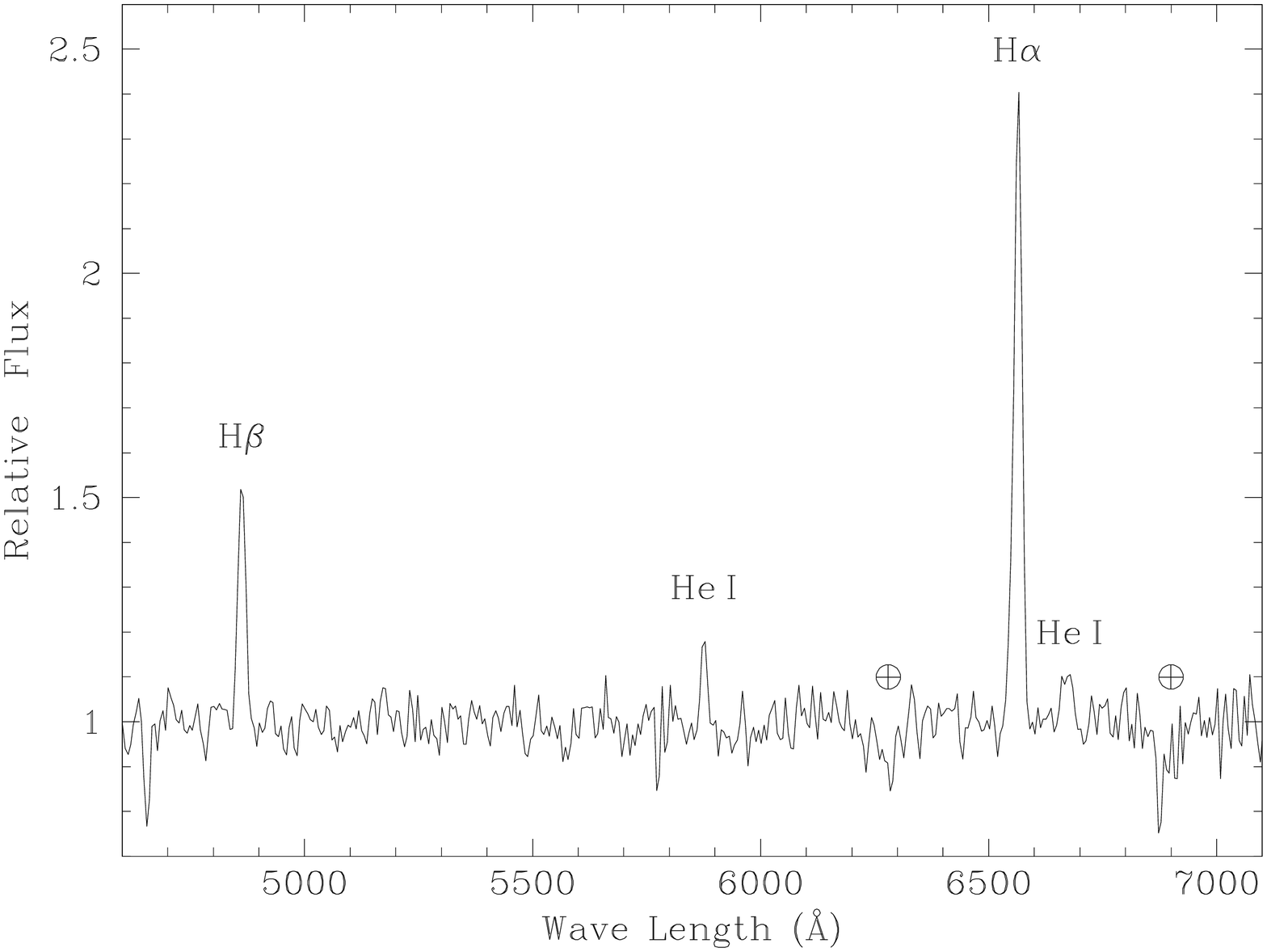}
  \end{center}
  \caption{Spectrum of RX And taken on 1997 February 5, during the
  decline phase from the third peak.  The Balmer and HeI lines are
  clearly seen in emission.}
  \label{fig:spec}
\end{figure}

    In the spectrum obtained on 1997 February 5 (figure \ref{fig:spec}),
when RX And was declining from the the third peak after recover from the
deep minimum and $\sim$1 mag above the deep quiescence, emission lines of
the Balmer lines, H$\alpha$ and H$\beta$, and HeI dominated.
The equivalent widths, FWHM and FWZI are listed in table \ref{tab:emission}.
The errors of FWHM and FWZI are 40 and 100 km s$^{-1}$, respectively.
The instrumental width corresponds to 540 km s$^{-1}$.
The H$\beta$ equivalent width is
$\sim10$\% larger than that in normal quiescence and on one day before an
outburst measured by \citet{kai88rxand}.  While most dwarf novae, including
RX And itself (cf. \cite{cla84rxandktper}; \cite{kai88rxand}) are known
to show reduced line strengths above quiescence, RX And in this stage
showed relatively strong emission lines.  This suggests that the
accretion disk had an extra emission source, which can be naturally
explained if the heating wave from the inner region had not reached
the outer region of the accretion disk and the accumulated matter during
the low quiescence was left from being accreted.  The large values of
FWHM and FWZI are also consistent with formation of these lines
in the outer disk.

\begin{table}
\caption{Widths of emission lines.}\label{tab:emission}
\begin{center}
\begin{tabular}{lccc}
\hline\hline
Line        & E.W.$^*$ & FHWM$^\dagger$ & FWZI$^\dagger$ \\
\hline
H$\alpha$   & $-$30 &  940  & 2650 \\
H$\beta$    & $-$10 & 1130  & 2630 \\
HeI 5876    & $-$3  &  670  & $\cdots$ \\
\hline
 \multicolumn{4}{l}{$^*$ Equivalent width in \AA.} \\
 \multicolumn{4}{l}{$^\dagger$ Unit in km s$^{-1}$.} \\
\end{tabular}
\end{center}
\end{table}

In conclusion, we interpret that RX And mimicked a long-period,
low-$\dot{M}$ dwarf nova in its recovery process from the deep minimum.

\subsection{Frequency of Deep Fadings}

   The reason why such deep fadings of RX And had not been recognized
before the 1996--1997 event was partly from the presence of a close
$V$=14.3 mag companion, which may have been confused with RX And when
the variable was exceptionally faint.  This hypothesis is strengthed
by the subsequent detections of (shorter) fading episodes occurring
1997 February to March and 2000 April to August.  During both fadings,
RX And was observed fainter than its ``nominal" quiescent magnitude
of 14.0.  Such fading episodes may have been identified as ``intervals
without outbursts" in the past visual record.  An examination of
the data in VSOLJ (Variable Star Observers League in Japan) and AFOEV
(Association Fran\c{c}aise des Observateurs d'Etoiles Variables) databases
since 1970 has revealed at least two distinct such instances in 1977
and 1986.  Although the frequency and periodicity should be confirmed
future observations, there seems to be a quasi-period of $\sim$10 yrs,
which is close to what is proposed for a VY Scl-type star, MV Lyr
(\cite{wen83mvlyr}; \cite{ros93mvlyr}).

\subsection{Relation to VY Scl-Type Stars}

   The present extended fading episode strongly mimics the temporary
fadings of VY Scl-type stars.  Observations have suggested a number of
similarities between the present phenomenon of RX And of VY Scl-type
fadings.  The presence of a similar periodicity between RX And and MV Lyr
may further suggest a common underlying mechanism for producing temporary
deep fadings, though the reason remains a mystery why there is no evidence
of a similar phenomenon in other Z Cam-type stars.

   Additional difference between the present fading of RX And and
VY Scl-type fadings can be seen in the initial part of the fading episode
(figure \ref{fig:overall}).  The disk instability model predicts
that a system should undergo dwarf nova outbursts (\cite{hon94v794aql},
\cite{kin98DI}) in response to the temporary reduction of $\dot{M}$.
This is what was indeed observed in RX And, in the early stage of
the fading.  VY Scl-type systems, on the contrary, show little such
evidence of dwarf nova-type outbursts (\cite{hon94v794aql};
\cite{gre98vyscl}).  \citet{lea99vyscl} explained that such a departure
of VY Scl-type stars from the disk-instability model is caused by
the irradiation from the hot white dwarf, which can suppress thermal
instability.  More recent discovery of unique short-period mini-outbursts
in a VY Scl-type star, V425 Cas \citep{kat01v425cas}, can be regarded
as another support to this explanation.  The detection of supersoft X-rays
from V751 Cyg in low state (\cite{gre99v751cyg}; \cite{gre00v751cygvsge})
and the indirect evidence in BZ Cam \citep{gre01bzcam}
also support this interpretation.  The far-UV observation by
\citet{sio01rxand} during the deep fading yielded a white dwarf temperature
of 34100$\pm$1000 K, which does not significantly differ from those
measured in other quiescent dwarf novae.  This observation provides
strong support for the interpretation that what phenomenologically
distinguishes temporary fadings of RX And from VY Scl-type fadings is the
effect of heating from the white dwarf.

\vskip 3mm

   We are grateful to many amateur observers for supplying their vital
visual and CCD estimates via VSNET.  We are grateful to staffs of
the VSOLJ and AFOEV for maintaining their databases publicly available.
We are grateful to the staffs of Okayama Astrophysical Observatory for
helping with our observation.
This work is partly supported by a grant-in-aid (13640239) from the
Japanese Ministry of Education, Culture, Sports, Science and Technology.

\end{document}